\documentclass[a4paper,12pt]{article}
\usepackage{amsmath,graphicx,amssymb}
\usepackage{authblk}
\usepackage{epsfig}

\makeatletter
\def\@maketitle{%
  \newpage
  \null
  \vskip 2em%
  \begin{center}%
  \let \footnote \thanks
    {\Large \bfseries \@title \par}%
    \vskip 2.0em%
    {\normalsize
      \lineskip .5em%
      \begin{tabular}[t]{c}%
        \@author
      \end{tabular}\par}%
    \vskip 1em%
    {\normalsize \@date}%
  \end{center}%
  \par
  \vskip 1.5em}
\makeatother

\providecommand{\keywords}[1]{{\textit{Keywords:}} #1}
\providecommand{\PACS}[1]{{\textit{PACS:}} #1}
\newcommand{\HG}{\hat{\Gamma}}
\newcommand{\be}{\begin{equation}}
\newcommand{\ee}{\end{equation}}
\newcommand{\ba}{\begin{array}}
\newcommand{\ea}{\end{array}}
\newcommand{\baa}{\begin{array}}
\newcommand{\eaa}{\end{array}}
\newcommand{\bea}{\begin{eqnarray}}
\newcommand{\eea}{\end{eqnarray}}
\newcommand{\half}{\frac{1}{2}}

\newcommand{\Tr} {{\rm Tr}}
\newcommand{\p}{p}
\newcommand{\q}{q}
\newcommand{\kb}{\bar{k}} 

\newcommand{\ET}{{\cal E}}

\newcommand{\llat}{\lambda_L}

\newcommand{\N}{L}
\newcommand{\lef}{\tilde l}
\newcommand{\ttheta}{\tilde \theta}

\begin{document}

\title{Volume independence for Yang-Mills fields on the twisted torus.}

\author[1]{{ Margarita Garc\'{\i}a P\'erez}
 \thanks{margarita.garcia@uam.es}}
\affil[1]{Instituto de F\'{\i}sica Te\'orica UAM-CSIC, E-28049--Madrid, Spain}

\author[1,2]{{ Antonio Gonz\'alez-Arroyo} \thanks{antonio.gonzalez-arroyo@uam.es}
}
\affil[2]{
Departamento de F\'{\i}sica Te\'orica, 
Universidad Aut\'onoma de Madrid, E-28049--Madrid, Spain }

\author[3]{{ Masanori Okawa} \thanks{okawa@sci.hiroshima-u.ac.jp}
}
\affil[3]{Graduate School of Science, Hiroshima University,
        Higashi-Hiroshima, Hiroshima 739-8526, Japan 
        }

\date{{ \it Dedicated to the memory of our friend and colleague Pierre van Baal}}

\maketitle

\begin{abstract}

We review some recent results related to the notion of {\it volume independence} in 
SU(N) Yang-Mills theories. The topic is discussed in the context of gauge theories living on 
a $d$-dimensional torus with twisted boundary conditions. 
After a brief introduction reviewing the formalism for introducing gauge fields on a torus, we discuss 
how {\it volume 
independence} arises in perturbation theory. We show how, for appropriately chosen twist tensors, 
perturbative results to 
all orders in the 't Hooft coupling depend on a specific combination of the rank of the gauge group ($N$) 
	and the periods of the torus ($l$) given by $l N^{2/d}$, for $d$ even.
We discuss the well-known relation to non-commutative field theories and address certain threats to {\it volume 
independence} associated to the occurrence of tachyonic instabilities at one-loop order. We end by presenting some
numerical results in 2+1 dimensions that extend these ideas to the 
non-perturbative domain.
\end{abstract}



\newpage

\section{Introduction}

This review will focus on a very intriguing aspect of Yang-Mills theories, the interplay between
gauge and volume degrees of freedom. The idea was originally formulated in the context of 't Hooft's large $N$ 
limit~\cite{largeN}. It was put forward by Eguchi and Kawai~\cite{EK}, who 
conjectured that gauge theories become volume independent in the limit of large number of colours.
This observation allowed to map Yang-Mills theories into matrix models consisting on $d$ 
matrices of infinite rank, where $d$ is the dimensionality of space-time. On a lattice formulation, they 
were simply represented by gauge links living on a {\it reduced} single-point, $d$-dimensional lattice.
Although the original proposal, coined as EK {\it reduction},
turned out not to be correct for $d>2$, several ways out were soon proposed like the {\it Quenched}
\cite{quenched} and the {\it Twisted}   
\cite{TEK11}-\cite{GonzalezArroyo:2012fx2} EK reductions. More recently, other alternatives have also been analyzed. 
They include the so called {\it continuum large $N$ reduction} \cite{NN1}-\cite{NN5} and other proposals where reduction is
enforced by the addition of adjoint fermions \cite{unsal11}-\cite{Lohmayer2}, or through modifications of 
the Yang-Mills action that include double trace deformations \cite{unsal21},\cite{unsal22}. 

In this work we will adopt a more general point of view and depart from the strict large $N$, reduced volume 
limit. For that purpose, we will consider finite $N$ gauge theories living on a finite $d$-dimensional box. 
This set-up was introduced long ago by 't Hooft as a way to define, in a gauge invariant way, electric and 
magnetic fluxes in gauge theories~\cite{thooft}. In that context, the size of the box $l$ introduces an
additional expansion parameter that sets the scale for the running of the coupling constant. Asymptotic 
freedom guarantees that perturbation theory holds for small $l$, while confinement should set-in in the infinite 
$l$ limit. Therefore, the size of the box becomes a tunable parameter that allows to control the onset of 
non-perturbative effects. A large number of works have exploited this idea, monitoring the volume dependence 
of physical quantities as a way to get an insight into the non-perturbative 
dynamics~\cite{thooft}-\cite{TONYREV}\footnote{This list of references is far from complete. It puts the emphasis on 
those involving twisted boundary conditions. A program along the same lines for periodic tori has been
developed by Pierre van Baal and collaborators. All the relevant references can be found in the 
review~\cite{vanbaal:2000zc} and in the book with Pierre van Baal's collected works~\cite{vanbaal:book}.}. 

The main idea we want to put forward in this review is that $N$ and $l$ are 
intertwined parameters. Under certain premises, they always appear on a specific combination determined by 
the number of compactified dimensions \cite{Unsal:2010qh1}-\cite{Perez:2013dra3}. 
This observation is what leads to the concept of {\it volume independence}. It implies a strong form of volume
reduction that holds at finite $N$ and allows to trade finite $l$ by finite $N$ effects without altering the 
dynamics.
In what follows we will focus on the case where the compact manifold is a $d$-dimensional torus endowed
with twisted boundary conditions. A limiting case of this set-up is Twisted Eguchi Kawai (TEK) reduction. 
We will show that, for irreducible
twist tensors and an even number of twisted compactified directions, this combination is given by
$\tilde l= l N^{2/d}$. Following Refs. \cite{Perez:2013dra1}-\cite{Perez:2013dra3}, we will also show that {\it volume independence} 
holds in perturbation 
theory to all orders in 't Hooft coupling and discuss if and when it extends to the non-perturbative domain. 

The review will try to be self-contained. We will start with a brief and general introduction to twisted boundary
conditions and the definition of gauge fields on a twisted torus. 
This will be followed by the derivation of the Feynman rules in this set-up and
a discussion of {\it volume independence} within perturbation theory. 
Section~\ref{s.beyond} raises some concerns towards the extension of these ideas beyond the perturbative regime.
They include the occurrence of tachyonic instabilities within the perturbative expansion~\cite{Guralnik:2002ru},
and of symmetry breaking in the TEK model~\cite{IO}-\cite{Azeyanagi}. We argue how both can be avoided by a 
judicious choice of parameters in the theory. We end in sec.~\ref{s.2+1} 
by presenting the results of an exploratory analysis in 2+1 dimensions that tests these ideas in the
non-perturbative domain by means of lattice simulations~\cite{Perez:2013dra1},
and conclude in sec. \ref{s.con} with a brief summary.

\section{Yang-Mills fields on a twisted torus}
\label{s.generalities}

This section will review the basic formalism for introducing SU($N$) gauge fields on a torus. 
We will focus on those aspects that are relevant for the discussion of {\it volume independence} 
in an even number of compactified dimensions~\cite{Perez:2013dra1}-\cite{Perez:2013dra3}, as well as for Twisted Eguchi Kawai 
reduction in the large $N$ limit of Yang-Mills theories~\cite{TEK11}-~\cite{GonzalezArroyo:2012fx2}. The reader 
interested in a more complete presentation to the field is referred to~\cite{TONYREV}.

We will start by considering a $d$-dimensional torus with periods $l_\mu$.
Non-compact extra dimensions can be easily incorporated into the formalism but will be neglected for the 
discussion in this section. Gauge connections in this base space are $N\times N$ traceless hermitian matrices 
satisfying the periodicity conditions~\cite{thooft}: 
\be
\label{A_per_cond}
A_\mu(x+l_\nu \hat\nu)= \Omega_\nu(x) A_\mu(x)\Omega^\dagger_\nu(x)+
i \, \Omega_\nu(x)
\partial_\mu\Omega^\dagger_\nu(x)\, .
\ee
The SU($N$) matrices $\Omega_\mu (x)$ are transition matrices characterizing the gauge bundle. They are 
subject to the consistency conditions:
\be
\Omega_\mu(x+l_\nu \hat \nu)\Omega_\nu(x)= Z_{\mu\nu}  \Omega_\nu(x+l_\mu \hat \mu)\Omega_\mu(x)\, ,
\ee
where $Z_{\mu\nu}= \exp \{2 \pi i n_{\mu \nu} / N\}$ is an element of the center of SU($N$), with  
$n_{\mu \nu}$ an antisymmetric tensor of integers defined modulo $N$. Under a gauge transformation, the 
pair $\{\Omega_\mu, A_\mu\}$ changes as:
\bea
A_\mu \rightarrow \Omega(x) A_\mu(x) \Omega^\dagger (x) + i \, \Omega(x) \partial_\mu \Omega^\dagger (x)\, , \\
\Omega_\mu (x) \rightarrow \Omega(x+l_\mu \hat \mu) \Omega_\mu (x)  \Omega^\dagger (x)\, ,
\eea 
but the integers $n_{\mu \nu}$ remain invariant, uniquely characterizing the bundle. This type of boundary 
conditions, introduced by 't Hooft in ~\cite{thooft}, are known as {\it twisted boundary conditions} 
and $n_{\mu \nu}$ as the twist tensor. 

One can make use of the gauge freedom to fix the value of the twist matrices $\Omega_\mu(x)$. 
In this review, we will focus on the analysis of twist tensors that allow for the choice of constant 
twist matrices $\Omega_\mu(x)=\Gamma_\mu$. They are known under the name of twist-eaters and satisfy:
\be
\label{eq.tweat}
\Gamma_\mu \Gamma_\nu = Z_{\mu\nu} \Gamma_\nu \Gamma_\mu\, .
\ee
Special relevance among those, play the so-called {\it irreducible twist tensors}, for which the solutions to 
Eq.~(\ref{eq.tweat}) are unique modulo similarity transformations (global gauge transformations) 
and multiplication by an element of 
$Z\!\!\!Z_{N}$~\cite{TONYREV}. It can be shown that, for irreducible twists, the number of inequivalent twist-eaters
is discrete and equal to $N^{(d-2)}$. Although to achieve our purposes it will not be necessary to discuss 
specific solutions, we point out that there is a general way to construct them in arbitrary number 
of dimensions \cite{baaltwe}, \cite{politw}.

Let us now consider the case of even number of compactified dimensions. In $d=2$, there 
is a unique twist tensor element $n_{12}$ and the twist is irreducible if $n_{12}$ and $N$ are coprime. 
In that case the
solution to Eq.~(\ref{eq.tweat}) is unique modulo similarity transformations. The SU($N$) $\Gamma_i$ matrices 
are traceless and verify the following conditions:
$$
\Gamma_i^N = \pm I\!\!\!I\, ,
$$
for $N$ odd or even respectively.

In four dimensions, the necessary and sufficient condition for the existence of solutions to Eq.~(\ref{eq.tweat}) 
is that the twist tensor satisfies $\kappa(n_{\mu \nu})  \equiv  \epsilon_{\mu \nu \rho \sigma} 
n_{\mu \nu} 
n_{\rho \sigma} /8 = 0\ ({\rm mod} \ N)$. This case is known as orthogonal twist. It is irreducible provided
the greatest common divisor of $N$, $n_{\mu \nu}$, and $\kappa(n_{\mu \nu})/N$ is equal to 1. As discussed above, 
for a given irreducible twist there are $N^2$ inequivalent solutions satisfying: $$\Gamma_i^N = I\!\!\!I \, .$$ 

For the discussion of volume independence, we will consider the set of twists given by
\be
\label{eq.twist}
n_{\mu\nu} =\epsilon_{\mu \nu} \, {k  N \over L }\, ,
\ee
with $k$ and $L$ integers, $L\equiv N^{2/d}$, and where 
\be
\epsilon_{\mu\nu} = 
\Theta(\nu-\mu)-\Theta(\mu-\nu)\, ,
\ee
with $\Theta$ the step function. If $k$ and $L$ are coprime these are irreducible
twists.  Let us consider now the set of $N\times N$ matrices:
\be
\label{eq.basis}
\HG(s) = \frac{1}{\sqrt{2N}}\, e^{i \alpha(s)} \, \Gamma_0^{s_0} \cdots \Gamma_{d-1}^{s_{d-1}}
\ee
where $s_\mu$ are integers. It can be shown that there are $N^2=L^d$ linearly independent such matrices. 
They constitute the algebra of twist-eaters which is isomorphic to the Lie algebra of U($N$)~\cite{EguchiNakayama}-\cite{TEK12}. 
In particular, for our choice of twist tensor, it can be shown that all $\HG(s)$ are traceless except 
for those satisfying $s_\mu=0$ (mod $L$), $\forall \mu$. The elements in the Lie algebra of SU($N$) can be hence
parametrized by the $L^d$ lattice of integers $(s_\mu)$, with $s_\mu = 0, \cdots, L-1$, excluding $s_\mu=0$, $\forall \mu$.
This will turn out to be useful below for solving the periodicity condition on the gauge fields.

In the formalism of constant twist matrices, the gauge potential has to satisfy the following boundary
conditions:
\be
A_\mu(x+ l_\nu \, \hat\nu)= \Gamma_\nu A_\mu(x)\Gamma^\dagger_\nu \ .
\ee
Notice that zero-action solutions (flat connections) with $A_\mu=0$ are compatible with these conditions. From
this observation, it is easy to determine the number of gauge-inequivalent flat connections. 
It suffices to take into account that $A_\mu=0$ is invariant under global gauge transformations, which, 
however, modify the twist matrices into: 
$\tilde \Gamma_\mu = \Omega \Gamma_\mu \Omega^\dagger$.
The number of inequivalent zero-action solutions is thus discrete and equal to the number of 
inequivalent twist-eaters~\cite{TONYREV}. The different solutions can be characterized by the 
value of non-zero Polyakov lines. On the twisted box they are defined as:
\be
\label{eq.pol}
{\cal P}(\gamma)\equiv \mathrm{Tr}\left( T\exp\{ -i g \int_\gamma dx_\mu A_\mu(x)\}
\, \Gamma_0^{\omega_0(\gamma)} \cdots \Gamma_{d-1}^{\omega_{d-1}(\gamma)} \right)
\, ,
\ee
where $\gamma$ is a closed curve on the d-torus and $\omega(\gamma)$ its corresponding winding number. 
The symbol $T\exp$ stands for the path-ordered exponential, where the order of matrix multiplication
follows left-to-right the order of the path. For a zero vector potential, the Polyakov lines are equal,
modulo a phase and the normalization, to $\Tr \, \HG(\omega(\gamma))$, with $\HG(\omega(\gamma))$ given 
by Eq.~(\ref{eq.basis}) with $s_\mu = \omega_\mu(\gamma)$.  As already mentioned, the only elements having
a non-zero trace are those for which $\omega_\mu(\gamma)=0$ (mod $L$), $\forall \mu$ (this will turn
out to be an essential ingredient in the discussion of
TEK reduction at weak coupling, as will be briefly discussed later on). The Polyakov loops
in those cases are phases in $Z_{N/L}$, giving rise to $N^{(d-2)}$ inequivalent solutions.

\section{Perturbation theory in the twisted box}
\label{s.perturbative}

The previous section provides all the necessary information to address perturbative calculations in the 
twisted $d$-dimensional box. The first step is to implement the boundary conditions on the vector potential.
This can be easily done if the gauge fields are expanded in terms of 
the Lie algebra basis provided by the $\HG$ matrices: 
\be
\label{eq.four}
A_\nu(x)=  {\cal N}\sum'_{p}   e^{i p \cdot x}\, \hat{A}_\nu(p)\, \HG(s(p)) \, ,
\ee
with ${\cal N}^{-2}= \prod_\mu l_\mu$.
The boundary conditions are automatically satisfied if the $p_\mu$ in this expression are quantized as:
\be
\label{eq.mom}
\p_\mu = \frac{2 \pi m_\mu}{\N \, l_\mu } \, ,
\ee
with $s_\mu(p)$ p-dependent integers, defined modulo $L$, given by: 
\be
s_\mu (p) = \tilde \epsilon_{\mu\nu} \, \bar k \, m_\nu \, ({\rm mod} \, L)\, .
\ee
Here, $\bar k$ is an integer defined through the
relation:
\be
\label{bark}
k \bar k = 1 \, ({\rm mod} \, L)\, ,
\ee
and $\tilde \epsilon_{\mu\nu}$ is an antisymmetric tensor satisfying:
\be
\sum_{\rho} \tilde \epsilon_{\mu\rho} \epsilon_{\rho \nu} = \delta_{\mu\nu} \, .
\ee
The expression in Eq. (\ref{eq.four}) can be naturally interpreted as a Fourier expansion in terms of momenta $p_\mu$.
One peculiarity of this expansion is that momentum appears 
quantized in units of $\N \, l_\mu$. In addition, the prime in the sum restricts the allowed set 
of momenta, excluding those with $m_\mu = 0$ (mod $L$), $\forall \mu$. This ensures that the $A_\mu$ 
field is traceless and naturally provides an infrared cut-off to the theory. 

To make the notation simpler, we will write in what follows $\HG(p)$ instead of $\HG(s(p))$. 
 The corresponding matrices are given by: 
\be
\HG(p) = \frac{1}{\sqrt{2N}}\, e^{i \alpha(p)} \, \Gamma_0^{s_0(p)} \cdots \Gamma_{d-1}^{s_{d-1}(p)} \, .
\ee
The phase factors, $\alpha(\p)$, can be chosen to satisfy the following commutation relations:
\be
\label{eq.comm}
[\hat \Gamma(p), \hat \Gamma (q)] = i \, F(p, q , -p-q) \, \hat \Gamma(p+q) \, ,
\ee
with  
\be
F(p,q,-p-q)= -\sqrt{\frac{2}{N}}  \, \sin\left(\frac{ \theta_{\mu \nu}}{2} \, p_\mu q_\nu
\right) \, ,
\ee
playing the role of the SU($N$) structure constants in this particular basis.
We have introduced the antisymmetric tensor $\theta_{\mu \nu}$ defined as: 
\be
\label{theta}
\theta_{\mu \nu} =    \frac{ \N^2  \, l_\mu \,  l_\nu} {4\pi^2} \times  \, \tilde
\epsilon_{\mu \nu} \, \tilde \theta  \, ,
\ee
where the angle $\tilde \theta \equiv 2 \pi \bar k / \N$. 

In order to perform the perturbative expansion, we must first fix a gauge. We will use a generalized 
covariant gauge with gauge parameter $\xi$. The gauge fixed Lagrangian density reads:
\be
{\cal L} = \half \Tr ( F_{\mu \nu}^2 )+ {1 \over \xi} \Tr (\partial_\mu A_\mu )^2
- 2 \Tr (\bar c \, \partial_\mu D^\mu c)\quad ,
\ee
with  $D_\mu \equiv \partial_\mu - i g A_\mu$, the covariant derivative, and $c$,
$\bar c$ the ghost fields.
Introducing now the Fourier expansion of $A_\mu(x)$, we arrive at the following expressions for the
gauge field propagator:
\be
\label{eq:prop}
P_{\mu \nu} (\p, \q) = {1 \over \p^2} \Big
(\delta_{\mu \nu} - (1-\xi) \ {\p_\mu \p_\nu \over \p^2}\Big) \  \delta(q+p) \quad ,
\ee
and the ghost fields:
\be 
P_g(\p,\q) =  - {1 \over \p^2}  \delta(\q+\p)\quad \, ,
\ee
with momenta quantized as in Eq.~(\ref{eq.mom}).

The Feynman rules are also easily derived. One only has to take into account that the SU(N) structure constants
$f_{a b c}$ have to be replaced by the momentum dependent functions $F(p,q,\tilde q)$, appearing in the commutation 
relations Eq.~(\ref{eq.comm}). The resulting expressions are very similar to their infinite volume counterparts, 
including:
\begin{itemize}
\item
A 3-gluon vertex: 
$$ {1 \over 3!} V^{(3)}_{\mu_1 \mu_2 \mu_3}(\p^{(1)}, \p^{(2)},\p^{(3)}) \Big (\prod_{i=1}^3 A_{\mu_i} (\p^{(i)})\Big )  \, \delta \Big (\sum_{i=1}^3 \p^{(i)}\Big )\, ,$$
with:
\bea
V^{(3)}_{\mu_1 \mu_2 \mu_3}(\p^{(1)}, \p^{(2)},\p^{(3)})&=& i g
 {\cal N}  F(\p^{(1)},\p^{(2)},\p^{(3)}) \times \\
\Big (( \p^{(3)}-\p^{(2)})_{\mu_1} \delta_{\mu_2 \mu_3}
   &+&  ( \p^{(1)}-\p^{(3)})_{\mu_2} \delta_{\mu_1 \mu_3}
   +  ( \p^{(2)}-\p^{(1)})_{\mu_3} \delta_{\mu_1 \mu_2}
\Big ) \, , \nonumber
\eea
\item
A 4-gluon vertex:
$$ {1 \over 4!}  V^{(4)}_{\mu_1 \mu_2 \mu_3\mu_4}(\p^{(1)}, \p^{(2)},\p^{(3)},\p^{(4)}) \Big (\prod_{i=1}^4 A_{\mu_i} (\p^{(i)})\Big )  \, \delta \Big (\sum_{i=1}^4  \p^{(i)}\Big )\, ,$$
with:
\bea
&& V^{(4)}_{\mu_1 \mu_2 \mu_3 \mu_4} (\p^{(1)}, \p^{(2)},\p^{(3)},\p^{(4)}) = - g^2   {\cal N}^2 \times  \\
\Big( &&F(\p^{(1)},\p^{(2)},-\p^{(1)}-\p^{(2)}) F(\p^{(3)},\p^{(4)},-\p^{(3)}-\p^{(4)}) (\delta_{\mu_1 \mu_3} \delta_{\mu_2 \mu_4}-\delta_{\mu_2 \mu_3} \delta_{\mu_1 \mu_4})  \nonumber\\
+&&F(\p^{(2)},\p^{(3)},-\p^{(2)}-\p^{(3)}) F(\p^{(4)},\p^{(1)},-\p^{(4)}-\p^{(1)})
(\delta_{\mu_2 \mu_4} \delta_{\mu_3 \mu_1}-\delta_{\mu_3 \mu_4}
\delta_{\mu_2 \mu_1})  \nonumber\\
+&&F(\p^{(1)},\p^{(3)},-\p^{(1)}-\p^{(3)})  F(\p^{(2)},\p^{(4)},-\p^{(2)}-\p^{(4)}) 
(\delta_{\mu_1 \mu_2} \delta_{\mu_3 \mu_4}-\delta_{\mu_3 \mu_2}
\delta_{\mu_1 \mu_4}) \Big )\nonumber \, .
\eea
\item
A ghost-gluon vertex:
\be
V^{(gh)} =   -i  g  {\cal N}  F(\p^{(1)},\p^{(2)},\p^{(3)}) \
\, \p_\mu^{(1)} \  \bar c (\p^{(1)}) A_\mu (\p^{(2)})
c(\p^{(3)})  \, \delta \Big (\sum_{i=1}^3 \p^{(i)}\Big ) \,  .
\ee
\end{itemize}

Using this rules, it is easy to derive for instance the one-loop correction to the propagator. 
In Feynman gauge ($\xi=1$), the formula for the two-point vertex function, obtained by resuming 
the Lippmann-Schwinger series, reads:
\be
{\bf \Gamma}_{\mu\nu}^{(2)} =   - p^2 \delta_{\mu\nu} + \Pi_{\mu \nu } (\p)\quad ,
\ee
where $\Pi_{\mu \nu }$ is the vacuum polarization tensor, given at one-loop by~\footnote{This expression corrects an error in Eq. (B.14) 
of Ref.~\cite{Perez:2013dra1}.}:
\bea
\label{eq.vacpol}
\Pi_{\mu \nu} (p) &=&
\half g^2 {\cal N}^2 \sum_{q}
F^2(\p,\q,-\p-\q) \, {1 \over \q^2 (\p+\q)^2} \times
\\
&&\Big\{ 4 \, \Big (\delta_{\mu \nu} p^2 -p_\mu p_\nu\Big ) + (d-2)  \, \Big ((\p_\mu + 2\q_\mu) 
(\p_\nu  + 2\q_\nu) -  2 \delta _{\mu \nu} q^2 \Big ) \Big \}
\nonumber \, .
\eea
It can be easily proven that $\Pi_{\mu \nu}$ fulfills the Ward identity ($\p_\mu \Pi_{\mu \nu}=0$) if the 
regulator of the momentum sums
preserves shift symmetry ($q_\mu \rightarrow q_\mu + p_\mu$). For completeness, let us mention that if non-compact 
directions are added to the base manifold one should include the corresponding momentum integrals: 
\be
\int_{-\infty}^{\infty} \frac {d q_\mu}{2 \pi}\, ,
\ee
and take into account that $d$ in Eq.~(\ref{eq.vacpol}) represents the total dimensionality of Euclidean space 
(including non-compact directions).

With this we have all the ingredients required for the discussion of {\it volume independence} within 
the perturbative set-up.

\section{Volume independence in perturbation theory}
\label{s.volumeind}

It this section we will use the perturbative expansion described above to analyze
the dependence of perturbative results on the rank of the gauge group and the periods of 
the twisted torus. We will show that, to all orders in perturbation theory and for fixed value of the angle 
$\ttheta = 2 \pi \bar k / \N$, these two parameters appear only through the combination  $\lef_\mu = \N l_\mu$.  
This is what we term {\it volume independence} at finite $N$($\equiv L^{d/2}$). It implies that different 
SU($N$) theories defined on torus manifolds with different periods become physically equivalent, at least at the 
perturbative level.

The first indication in this respect comes from the momentum quantization rule
Eq.~(\ref{eq.mom}). Disregarding the fact that zero momentum (mod $L$) is not allowed in the twisted box, momentum is 
quantized as if the theory was defined on a torus with extended periods $\lef_\mu$. 
We have also seen that all vertices in perturbation theory are proportional to the factor:
\be
g \, {\cal N} F(p,q,-p-q) = - \sqrt{\frac{2\lambda} {\prod_\mu \lef_\mu}}\,
\, \sin\Big (\frac{\theta_{\mu \nu}}{2} \, p_\mu  q_\nu\Big ) \, ,
\ee
with
\be
\label{eq.noncom}
\theta_{\mu \nu} = \frac{ \lef_\mu \,  \lef_\nu} {4\pi^2} \times  \, \tilde
\epsilon_{\mu \nu} \, \tilde \theta  \, .
\ee
preserving, for fixed $\tilde \theta =2\pi \bar k /\N$, the dependence on $\lef$ to all orders. 

Let us now examine some of the consequences of this particular perturbative expansion. The first one is the
well known relation to non-commutative gauge theories \cite{douglasnekrasov}. It is derived from the fact that the 
coefficients $\hat A_\mu(p)$ of the Fourier expansion in Eq.~(\ref{eq.four}) 
are pure complex numbers. They give rise to a propagator without colour 
degrees of freedom, as the one corresponding to a U(1) gauge theory. It is though a peculiar U(1) theory with 
momentum dependent phases, proportional to $\sin(\theta_{\mu \nu} p_\mu q_\nu /2)$, entering the vertices.
This relation is the perturbative manifestation of Morita duality~\cite{GAKA}-\cite{Ambjorn3}, stating 
that the SU(N) twisted theory is physically equivalent to a non-commutative U(1) gauge theory defined on a periodic 
torus with periods $\lef_\mu$ and non-commutativity parameter $\theta_{\mu \nu}$. 
Strictly speaking, the Morita mapping applies to the U(N) gauge theory, including momentum modes in 
the original torus that are zero (mod $L$). On the non-commutative side, they give rise to photon modes 
with momenta quantized in units of 
$2 \pi /l_\mu$. Due to the form of the structure constants, 
these modes decouple and do not interfere with the duality. Suppressing them is, however, essential to avoid the 
existence of infrared divergences in the original torus and, as we will see, to prevent the appearance of tachyonic 
instabilities in the theory.

The combined $N$ and $l$ dependence of the perturbative expansion has far-reaching consequences. 
{\it Volume independence} also implies an equivalence between different SU($N$) commutative gauge theories, provided $\lef$ and $\ttheta$ 
are kept fixed. To be strict, however, we have to point out one possible caveat. It is derived from the
impossibility to rigorously keep $\ttheta$ fixed as $N$ changes. This is so 
because $\ttheta/(2\pi)$ is a rational number with coprime rational factors $\bar k$ and $N^{2/d}$. 
For volume independence to hold, one has to assume 
that all gauge invariant quantities depend smoothly on
$\ttheta$.  This issue is difficult to settle in general terms and has been analyzed by several authors in the
context of non-commutative field theories - see e.g. the discussions related to
the application of Morita duality at irrational values of $\ttheta$ in Ref. \cite{AlvarezGaume:2001tv}.
We will come back to this important point in sec.~\ref{s.inst} when discussing the appearance of tachyonic 
instabilities in perturbation theory 
following Ref. \cite{Guralnik:2002ru}.
 
Let us finally mention that a particular case of this equivalence is Twisted Eguchi Kawai (TEK) reduction. 
It corresponds to a discretized version
of large N Yang-Mills theory on a periodic lattice with a single lattice site. In our context this would correspond to a 
limit in which the torus periods have the length of one lattice spacing, giving $\tilde l = \N a$. For TEK reduction, 
the large $N$ limit is taken first at fixed value of the lattice spacing.  After that, the continuum limit is approached, 
driven by the large $N$ beta function. The resulting theory is claimed to be equivalent to an infinite volume, 
SU$(\infty)$ Yang-Mills theory in the continuum. The first proofs of reduction \cite{EK} were based on the equivalence 
of the Schwinger-Dyson equations satisfied by the Wilson loop observables in the original theory 
and those of the reduced theory. The proof relied on large $N$ factorization and required certain symmetries of the 
theory to be preserved. In particular, it was essential to have zero expectation value for open Wilson lines 
(Polyakov loops in the reduced theory). It was soon realized that this was not the case for a strictly periodic lattice. 
The problem appeared already at the perturbative level, since the allowed flat connections did
not satisfy this condition. Very early after this, two of the present authors pointed out a solution based 
on the introduction of twisted boundary conditions~\cite{TEK11},\cite{TEK12}.
For the type of irreducible twists discussed here, we know that the allowed flat connections
have zero Polyakov loops except when the winding number is $0 \, ({\rm mod} \, \N)$. 
The symmetry requirement is thus fulfilled at weak coupling 
except for loops of length $\N$ ($\rightarrow \infty $ in the large $N$ limit). 
In addition, Refs.~\cite{TEK11},\cite{TEK12} provided an alternative derivation of reduction based on perturbation theory on a 
twisted torus, along the lines presented here.
The non-trivial Feynman rules giving rise to non-commutative dynamics were also anticipated in \cite{TEK12,GAKA}, 
preceding by many years the introduction of non-commutative field theories. 
As a matter of fact, TEK models have been used in the past \cite{Ambjorn1}-\cite{Ambjorn3}
as a regularized version of non-commutative gauge theories with non-commutativity parameter:
\be
\theta_{\mu \nu}^{\rm TEK} =    \frac{ \N^2  \, a^2} {4\pi^2} \times  \, \tilde
\epsilon_{\mu \nu} \, \tilde \theta  \, .
\ee
In that context, most of the results analyzed so far in the literature were concerned with the case of  $\tilde \theta $ scaling like
$1/\N$ in the large $N$ limit. This gives rise to a continuum non-commutative limit only in the so called {\it double 
scaling limit} where $\N a^2$ is kept fixed as the large $\N$, $a\rightarrow 0$, limit is taken. Following the discussion 
above, we will be analyzing instead the limit in which the large $N$ limit is taken by sending 
$\lef \longrightarrow \infty$, while keeping $\tilde \theta $ fixed. 

\section{Going beyond perturbation theory}
\label{s.beyond}

We have shown how {\it volume independence} works at a perturbative level. 
Whether it is also preserved non-perturbatively is an issue much more difficult to settle.  
In this section, we will discuss several reasons for concern that have been raised 
in the literature. They include the appearance of instabilities of the perturbative 
vacuum  \cite{Gomis:2000pf}-\cite{Bietenholz:2006cz2},
and of symmetry breaking in the TEK model at large values of $N$ \cite{IO}-\cite{Azeyanagi}. 
Together with a generic discussion of 
the problems, we will show how to prevent them by appropriately scaling 
the parameters of the theory~\cite{TEK21}, 
\cite{Perez:2013dra1}.

\subsection{Tachyonic instabilities}
\label{s.inst}

Soon after the appearance of non-commutative theories in the string theory literature, it was 
realized that these theories lead to problems at a perturbative level. 
In particular, it was shown that certain low momentum modes can become tachyonic and render the 
perturbative vacuum unstable \cite{Gomis:2000pf}-\cite{Armoni:2003va}. 
Using the mapping between commutative and non-commutative theories just described, this could apply as well to
the commutative case on the twisted torus~\cite{Guralnik:2002ru}. The commutative theory would, of course, never 
become tachyonic, but the presence of these modes was argued to induce a breaking of translational invariance which was 
indeed detected through non-perturbative lattice simulations in certain models \cite{Bietenholz:2006cz1,Bietenholz:2006cz2}.
In this section we will present the set-up leading to these conclusions and we will argue that the 
tachyonic behaviour can be avoided through a judicious choice of parameters in the theory ($N$, $k$, and $\kb$),  
while still preserving volume independence.

To set the stage, we will discuss how tachyonic modes appear in the SU($N$) gauge theory for
two twisted compact directions, following Refs. \cite{Perez:2013dra1}, \cite{Guralnik:2002ru}. The base manifold 
we will be considering is $T^2 \times R^{d-2}$. We will take one of the infinite directions to play 
the role of Euclidean time\footnote{For $d=4$, coordinates in this space will be labelled as $(x_0,x_1,x_2,x_3)$, 
with $x_1$, $x_2$ the twisted directions and $x_0$, $x_3$ the non-compact ones}. This allows to define a spectrum of 
states in a Hamiltonian set up. We will be 
concerned in particular with states of non-zero electric flux \cite{thooft}-\cite{Pierreth2}. 
In the twisted box, electric flux arises as a quantum number associated to the action of the so called {\it
singular gauge transformations}. These are SU(N), time independent, transformations that satisfy the generalized 
periodicity conditions:
\be
\Omega_{[\vec K]} (\vec x+K_i \hat{\imath}) = e^{2 \pi i K_i \over N} \, \Gamma_i \Omega_{[\vec K]} (\vec x) \Gamma_i^\dagger\, .
\ee
For $\vec K \ne \vec 0$, they are symmetries of the action which, however, do not correspond to gauge transformations since they modify 
the Polyakov loops by an element of the center of the group. 
Let us label the space of SU($N$) matrices satisfying the
previous equation by ${\cal G}(\vec K)$. A particular representative for given $\vec K$ is the constant matrix defined as:
\be
\label{omegaK}
\Omega_{[\vec K]} = \Gamma_1^{\kb K_2} \Gamma_2^{-\kb K_1} \, .
\ee
The representations of the quotient group:
\be
\left(\cup_{\vec{K}} \, {\cal G}(\vec{K})\right)/{\cal G}(\vec{0}) \sim Z_N^2 \, ,
\ee 
are labelled by the electric flux vector $\vec e$ (defined modulo $N$). There are thus $N^2$ electric flux 
sectors and the Hamiltonian can be independently diagonalized in each of them.   
Under the operator that implements these transformations, the elements of the Hilbert space carrying electric flux 
$\vec e$ transform as: 
\be
{\bf U} (\Omega_{[\vec K]}) |\Psi(A)> = e^{i 2 \pi {\vec e\cdot \vec K \over N}} |\Psi(A)> \, .
\ee
They can be constructed in terms of Polyakov loop operators defined in Eq.~(\ref{eq.pol}). 
Using the transformation properties under $\Omega_{[\vec K]}$ in Eq. (\ref{omegaK}), it is easy to see that these 
gauge invariant operators carry electric flux given by their winding number (modulo $N$). 
In addition, they satisfy non-trivial boundary conditions along the compact twisted directions. This enforces a relation
between the electric flux and the momenta, appearing in the Fourier decomposition of the operators, given by:
\be
e_i = {\lef\over 2\pi } \, \epsilon_{ij} \, \bar k p_j   \, ({\rm mod} \, N), \,  {\rm for} \, \, i=1,\, 2\, .
\ee
In Ref.~\cite{Perez:2013dra1} we have derived, in perturbation theory, the energy spectrum of these states using several
alternative methods including the Hamiltonian quantization of the system in the $A_0=0$ gauge, and the Euclidean approach,
both on the lattice and in dimensional regularization. They all give consistent results. For concreteness, we will summarize 
here how to proceed in dimensional regularization.

The electric flux spectrum can be extracted in a gauge invariant way from the exponential decay at large time of Polyakov
loop correlators of a given winding number $\vec e$. To one-loop order this turns out to be proportional to the 
correlator of two transverse gluon fields with minimal non-zero momentum in the twisted ($i=1,2$) directions given by: 
\be
p_i = - {2\pi\over \tilde l } \, (\epsilon_{ij}\,   k e_j   \, {\rm mod} \, N) \equiv  {2\pi n_i \over \tilde l } \, .
\ee
Therefore, the energy spectrum can be derived from the poles of the gluon propagator for transverse gluons with 
$p_0= i {\cal E}$ (setting $p_3=0$, in the $d=4$ case). At zeroth order in perturbation theory the mass of the states 
within each electric flux sector is determined by the minimal momentum in that sector, giving ${\cal E} = |\vec p|$. 

The first correction in perturbation theory can be derived from the formula of the inverse propagator in Eq. 
(\ref{eq.vacpol}). For a certain transverse polarization $\varepsilon$ the resulting energy satisfies the following dispersion 
relation:
\be
{\cal E}^2 (\vec p ) = \vec {p}^{\, 2} + g^2 \delta {\cal E}^2(\vec p ) = \vec p^{\, 2}  - {\sum_\mu \varepsilon_\mu \Pi_{\mu \nu}^{\rm on-shell} 
\varepsilon_\nu \over \sum_\mu \varepsilon_\mu^2} \, . 
\ee
with $\Pi_{\mu \nu}$ evaluated for tree-level on-shell momenta with $p^2= 0$.
In 2+1 dimensions there is only one transverse polarization corresponding to $\varepsilon \propto (0,p_2,-p_1)$. 
In 2+2 dimensions, there are instead two which, for the momentum considered, correspond to: 
$\varepsilon^{(1)} \propto (0,p_2,-p_1,0 )$, and $\varepsilon^{(2)} = (0,0,0,1)$. The energy correction for polarization
$\varepsilon^{(2)}$ is given by $-\Pi_{33}$. For gluons polarized along the directions of the twisted torus, one can use 
the Ward identity to rewrite the self-energy correction in a simpler form, giving:
\be
{\cal E}^2 (\vec p ) = \vec p^{\, 2}  - \sum_{\mu=0}^2 \Pi_{\mu \mu}^{\rm on-shell}  \, ,
\ee
arriving at the simple expression:
\be
{\cal E}^2 (\vec p ) = \vec p^{\, 2}  + {2(d-2) \, \lambda  \over \tilde l_1 \tilde l_2} \int {d^{(d-2)} 
q \over (2\pi)^{(d-2)}}\, 
\sum_{q} \sin^2 \left(\frac{ \theta_{i j}}{2} \, p_i q_j \right) \, \Big ({d-2 \over \q^2 }  + \delta_{d,4}  \, {2 q_3^2 -q^2   
\over (p+q)^2 q^2 } \Big )\, .
\ee
We will simplify the analysis by setting $\lef_1 = \lef_2=\lef$. Rescale now the loop-momentum in all directions
to make it dimensionless:  $q_\mu = 2 \pi \hat q_\mu /\lef$. In the compact directions we take 
$\hat q_i \equiv m_i \in Z\!\!\!Z$. This allows to factorize out all the dependence in dimensionful quantities:
\be
{\cal E}^2 (\vec p ) = \vec p^{\, 2}  + {(d-2) \, \lambda  \over 2 \pi^2 \lef^{(d-2)}} \int d^{(d-2)}\hat q  \!
\sum_{\vec m} \sin^2 \left({\ttheta \over 2} \tilde \epsilon_{i j} \hat p_i \hat q_j \right) \! \Big ( {(d-2) \over \hat q^2 } + \delta_{d,4}   
{2 \hat q_3^2 - \hat q^2   \over (\hat p+\hat q)^2 \hat q^2 } \Big ) \, .
\ee
In reference \cite{Perez:2013dra1}, we have worked out in detail the $d=3$ case. After performing the integral in $q_0$,
the full expression can be rewritten in terms of Jacobi $\theta_3$ functions \cite{Tata}:
\be
\theta_3(z,it) = \sum_{m \in {\bf Z}} \exp\{-t \pi m^2 + 2 \pi i m z\}\, .
\ee
The final result is quite compact. Introducing the function:
\be
G ({\vec z} ) = -{1 \over 16 \pi^2} \int_0^\infty {dt \over \sqrt{ t}} \,
\Big(\theta_3^2(0,it) -\theta_3(z_1,i t) \, \theta_3(z_2,i t)- {1 \over t}\Big)\, ,
\label{eq:Gtheta}
\ee
it can be written as follows:
\be
\label{elenergform}
{\ET^2(\vec e)  \over \lambda^2} =  {|\vec n|^2 \over 4 x^2}
- {1 \over x } \, G  \Big({\tilde \theta \vec{\tilde n} \over 2 \pi}\Big)\, ,
\ee
where $\vec n$ is the minimal momentum in each electric flux sector:
\be
- {N \over 2} < (n_i = -k \, \epsilon_{ij} e_j \, ({\rm mod} N)) < {N \over 2} \, ,
\ee
and $\tilde n_i = \epsilon_{ij} n_j$.
Recalling that $\tilde \theta = 2 \pi \kb/N$, the argument of the function $G$ turns out to be the
electric flux divided by $N$. Notice that
we have written the expression for the energy in terms of the variable $4 \pi x= \lambda \lef $.
This is quite natural, since in 2+1 dimensions $\lambda$ has
dimension of energy, and appears as the natural unit. In the 2+1 dimensional case, it
is indeed $x$ the variable that controls the size of the one-loop correction. 

\begin{figure}
\centerline{
\psfig{file=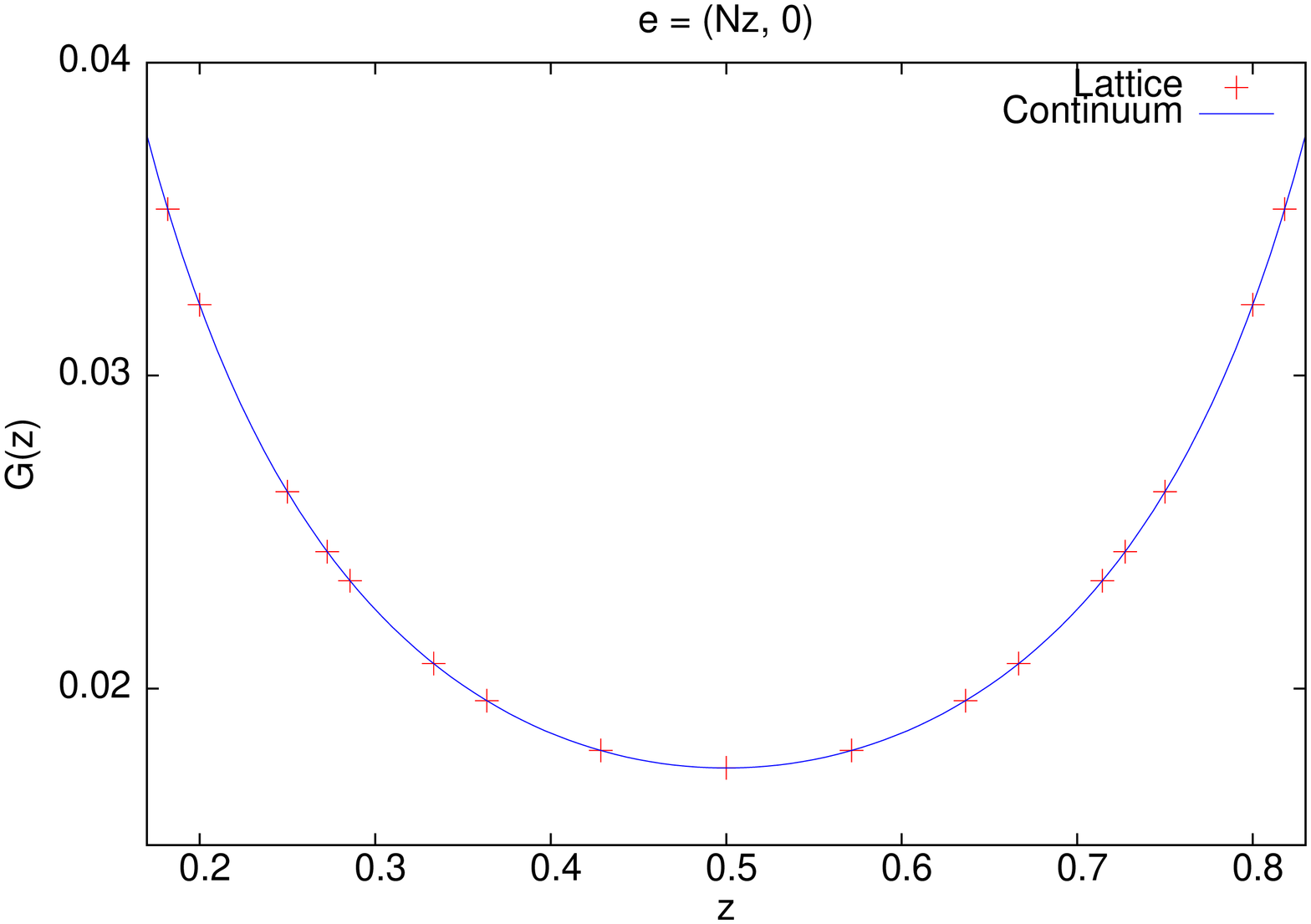,angle=-90,width=7cm}
\psfig{file=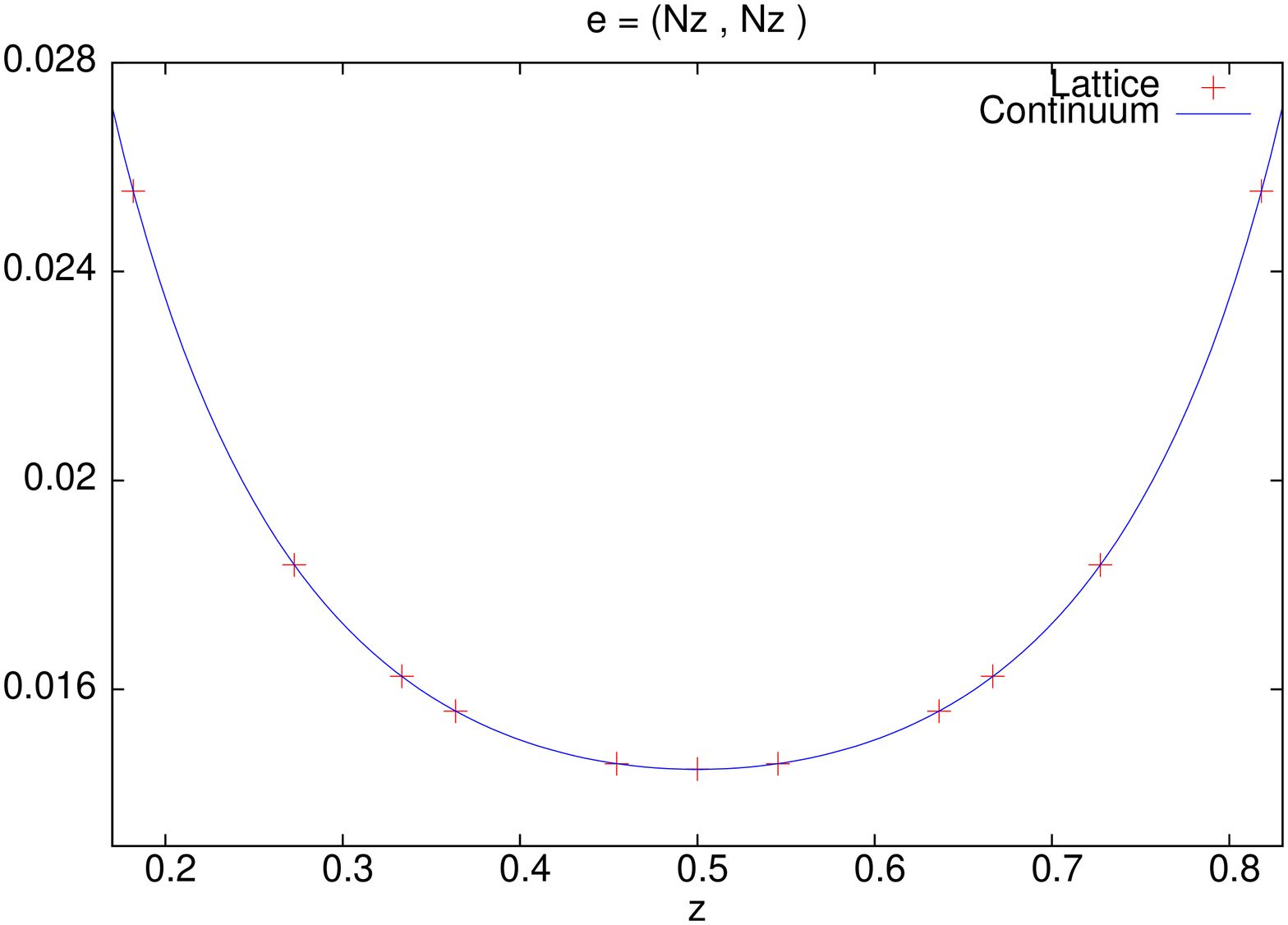,angle=-90,width=7cm}}
\caption{
We show, for electric flux $\vec e =(N z, 0)$ (left), and $\vec e =(N z,N z)$ (right), the function $G(z)$ that
gives the one-loop correction to the energy of electric flux $\vec e$,
through Eq.~(\ref{elenergform}). The blue line corresponds to the continuum expression Eq.~(\ref{eq:Gtheta}),
while the red points are derived using a lattice regularization in the calculation of the self-energy.}
\label{fig.theta}
\end{figure}

We have now all the required ingredients to discuss whether the perturbative vacuum becomes unstable at the
one-loop level. Notice that, since the first term in the dispersion relation is just the momentum squared,
it is natural to interpret the correction as the mass squared. However,
the function $G(\vec{z})$ is positive, giving rise to a negative mass squared
contribution.  This does not necessarily imply a negative energy squared (a tachyon). At tree level 
the theory has a mass gap, and for arbitrarily small coupling it is stable. However, as the coupling increases,
a tachyonic instability seems to unavoidably appear at a critical value of $x$ given by:
\be
\label{eq.inst}
x_c = {|\vec n|^2 \over 4 G  (\ttheta \vec{\tilde n} /(2 \pi))}\, ,
\ee
The question is then whether this occurs at sufficiently small coupling
for perturbation theory to be reliable.
In order to analyze this, let us first look at the structure of the function $G(\vec{z})$.  
To illustrate the $\vec{z}$ dependence, we display in fig. \ref{fig.theta} 
the function $G (\vec z)$ for two different cases: $\vec z= (z,0)$, and $\vec z=  (z,z)$.
It is positive and it strongly peaks at $z$ close to $0$ and $1$. 
From the analytic formula it can be shown that it indeed diverges for  $\vec z=
\vec 0$  (mod 1). It is relatively simple to compute the behaviour
close to the singularity by using the duality  relations of the $\theta_3 $ function:
\be
\theta_3 (z,it) = {1 \over \sqrt{t}} \, e^{-{\pi z^2 \over t}} \sum_{k
\in {\bf Z}} \exp\{-{\pi k^2 \over t} + {2 \pi z k \over t}\}\, .
\ee
Focusing on the integral producing the divergence,
\be
\label{Gasymp}
\int_0^1 {dt \over t^{3/2}} \ \exp \Big (-{\pi  |\vec z|^2\over t }\Big  )
=  \, {1 \over \ |\vec z| } + {\rm regular \ terms} \, ,
\ee
we derive:
\be
{\ET^2(\vec e)  \over \lambda^2} =  { |\vec n|^2 \over 4 x^2}
- {1 \over  8 \pi x \, |\ttheta \vec n|  }  \quad ,
\ee
valid for $|\ttheta \vec n|\rightarrow 0 $. Notice that, due to the periodicity properties of $G(z)$,
there is also a divergence whenever $\ttheta \vec n = \vec 0$  (mod $2\pi$). 

Using this result, one derives that the instability appears at:
\be
x_c = 2 \pi |\vec n|^2 |\ttheta \vec n| \equiv {4 \pi^2 |\vec n|^2 |\vec e| \over N}\, .
\ee
Taking in this formula $|\vec n|$ and $|\vec e|$ small, seems to unavoidably imply that
$x_c \rightarrow 0$ in the large $N$ limit. This would thus break volume independence, introducing a difference
between the large volume and the large $N$ behaviour.  The argument, first introduced in Ref. \cite{Guralnik:2002ru}, 
can be generalized to the case with two non-compact dimensions. In the limit $|\ttheta \vec n|\rightarrow 0$, 
the general formula reads \cite{Guralnik:2002ru}:
\be
\label{instability}
{\ET^2(\vec e) } =  { 4 \pi^2 |\vec n|^2 \over \lef^2}
- {\lambda 2^{(d-3)} (d-2) \Gamma(d/ 2)  \over  (\pi)^{2-d/2}} \, \,   
\Big( \lef \, |\tilde \theta \vec n|\Big)^{(2-d)}   \, .
\ee
Instabilities for $|\ttheta \vec n|\rightarrow 0$ thus appear at a critical coupling:
\be
\lambda_c \, \lef^{\, (4-d)}_c  \propto  |\vec n|^2 |\ttheta \vec  n|^{(d-2)} \, .  
\ee
In order to avoid a small $\lambda_c$, one has to require that $|\vec n|^2$ grows at least as 
$|\ttheta \vec n|^{(2-d)}$, when  $|\ttheta \vec  n|$ becomes small.
This condition sets limits on the allowed values of $k$ and its conjugate $\kb$. To see that, 
let us consider two extreme cases: 
\begin{itemize}
\item
The lowest non-zero electric flux  $|\vec e|=1$ corresponds to  $|\ttheta \vec n|=2\pi/N$ and $|\vec n|=k$. This
implies that $k$ has to scale in the large $N$ limit at least as $\sim N^{(d-2)/2}$. 
\item
The lowest non-zero momentum $|\vec n|=1$ corresponds to $|\ttheta \vec n|=\ttheta$. This would lead to problems 
unless $\ttheta > \ttheta_c$ in the large $N$ limit.
\end{itemize}
Summarizing, both $k$ and $\kb$ have to be scaled with $N$ as one takes the large $N$ limit.
These are necessary requirements. However, they might not be sufficient to guarantee stability.  
Some counter-examples were for instance provided in Ref.~\cite{Guralnik:2002ru}.  
To analyze the generic case, we will consider a sequence of SU($N$) theories with fixed $\lef= Nl$.
Take $N=Q \, N_0 - b$,  and  $\kb = Q \, \kb_0 + a $, with $N$, $N_0$, $\kb$, and $\kb_0$ prime numbers. For this set:
\be
\ttheta = 2 \pi \, { \kb_0 + a/Q \over  N_0 - b/Q }\, ,
\ee 
where $a$, $b$, and $Q$ are integers such that: $a <\!< \kb_0 Q$, and $b<\!< N_0 Q$, 
giving a value of $\ttheta$ approximately equal to $\ttheta_0 = 2 \pi \kb_0 / N_0 > \ttheta_c$.
For given $\ttheta_0$ and fixed $\lef$ this sequence should provide a set of smoothly related SU($N$)
theories, if volume independence holds. 
What happens if one approaches now the large $N$ limit by taking $Q$ large at fixed $\kb_0$, $N_0$? 
Volume independence would be broken
if any of the pairs $(|\vec n|, |\ttheta \vec n|)$ develops a tachyonic behaviour for large $Q$.
In this instance, there is a specific non-minimal momentum that can become problematic. It is $\vec n = (N_0, 0)$, 
which has an associated  value of $|\ttheta \vec  n|$ given by:
\be
|\ttheta \vec  n | =  {2 \pi N_0 \over N} \, |a + \ttheta_0 \, b| \, , 
\ee  
this gives
\be
\lambda_c \lef^{4-d} \propto  N_0^d \, \Big ({|a + \ttheta_0 \, b|\over N}\Big )^{(d-2)}\, .
\ee
It is clear that $\lambda_c$ tends to zero if the large $N$ limit is taken with fixed $a$ and $b$.
However, if $|a + \ttheta_0 \, b|$ is scaled with $N$ we can safely keep $\lambda_c$ away from the domain of 
reliability of perturbation theory. This can be done while still keeping the bounds on $a$ and $b$ that
guarantee an almost constant value of $\ttheta$, and a smooth dependence of the electric flux spectrum on
$\ttheta$.

\subsection{Symmetry breaking in the TEK model}

We will analyze now the limiting case of TEK reduction and discuss certain issues that arise
due to spontaneous symmetry breaking at a non-perturbative level.
We have already mentioned that reduction in the TEK model relies on the hypothesis that the reduced model 
respects the $Z\!\!\!Z_\N^d$ symmetry of the large volume theory. This is certainly the case in 
the weak coupling limit since the twist-eaters, for appropriate twist choices, 
respect the symmetry. However, the symmetry 
could be broken by non-perturbative effects. Indeed, simulations performed with the choice of twist originally 
proposed in Ref. \cite{TEK11} ($k=1$) showed a pattern of symmetry breaking at intermediate couplings 
\cite{IO}-\cite{Azeyanagi}.
The authors of Ref. \cite{Teper:2006sp} suggested that the origin of the symmetry breaking could be due
to other extrema of the TEK action functional known as fluxons \cite{vanBaal:1983eq}. They correspond to solutions of the 
equations of motion that satisfy consistency conditions given by:
\be
\label{eq.fluxons}
\Gamma'_\mu \Gamma'_\nu = e^{2 \pi i n'_{\mu \nu  }\over N}  \Gamma'_\nu \Gamma'_\mu\, .
\ee
with a twist tensor $n'_{\mu \nu  }$ different from the one characterizing the theory. 
These fluxons can have open paths with non-zero traces and induce $Z\!\!\!Z_\N^d$ symmetry breaking.
A extreme case is that of singular torons \cite{vanBaal:1983eq}, having $U_\mu  = z_\mu I\!\!I$. In that
case, the symmetry breaks down completely and all paths have non-zero trace. Since these configurations
have non-zero action, they are suppressed at weak coupling but  they
could dominate the partition function at intermediate couplings if entropy overcomes the difference 
in action with respect to the vacuum \cite{Teper:2006sp}.  We will reproduce here the discussion by two of the present authors
presented in Ref.\cite{TEK21}, and argue that an appropriate choice of twist can prevent this from happening.
Incidentally, let us point out that the criteria to avoid $Z\!\!\!Z_\N^d$ symmetry breaking coincide with the ones presented
in the previous subsection to prevent the appearance of tachyonic instabilities.

The action in the TEK reduced model is given by:
\be
S = N b \sum_{\mu \ne \nu} \Big (N - e^{i {2 \pi k \over \N} \, \epsilon_{\mu \nu } } \, \, 
\Tr (U_\mu U_\nu U_\mu^\dagger U_\nu^\dagger) \Big )
\ee
where $b$ is the inverse of the lattice bare 't Hooft coupling.
A singular toron, with $U_\mu  = z_\mu I\!\!I$,  has a difference in action with respect to a 
twist-eater given by:
\be
\Delta S = 2 d (d-1) \, b L^d \sin^2 \Big ({\pi k \over \N} \Big ) \, ,
\ee
where $d$ is the number of dimensions.  If the 
large $\N$ limit is taken at fixed $b$ and $k$, the difference in action grows as 
$L^{(d-2)}$ which 
can be overcome by an entropy growing as $L^d$ (given by the number of degrees of freedom in the system). 
Choosing instead a value of $k$ that scales as $\N$ would solve this problem. 
We stress that this criteria is one of those required to avoid the occurrence of 
tachyonic instabilities in the cases discussed in the previous subsection. Although the relation between fluxons and
tachyonic instability is not clear, in Ref~\cite{Guralnik:2002ru} it has been 
argued that the latter leads also to non-zero Polyakov loop expectation values and 
translational symmetry breaking, similar to the effects induced by fluxons. 

Ref. \cite{TEK21} also discusses possible more dangerous cases in which the entropy of the singular toron grows as 
$L^d \log(L)$ as suggested in \cite{zeromom}. Quantum fluctuations around these solutions
give an action for the singular torons \cite{vanBaal:1983eq}:
\be
S= d (d-1) \, {b \over 2}  \, \Big \{ 4 L^d \sin^2 \Big ({\pi k \over \N} \Big )
+  L^{d\over 2} \cos \Big ({2 \pi k \over \N} \Big )
\Tr(F_{\mu \nu}^2) \Big \} \, ,
\ee
showing that, for $k/\N > 1/4$, they become unstable and decay into twist-eaters, 
representing no longer a menace for TEK reduction. 

Although a formal proof of reduction away from
the weak coupling region is still lacking, the authors of Ref. \cite{TEK21} have performed a series of
detailed numerical studies \cite{TEK21}-\cite{GonzalezArroyo:2012fx2}, going to much larger values on $N$ than 
those previously explored in the literature. For values of $k$  satisfying $k/\N > 1/9$, with 
$\bar k/\N$ finite in the large $N$ limit, they have seen no evidence of symmetry breaking 
for values of $N$ up to $N=1369$. 

\section{Non-perturbative results in 2+1 SU(N) Yang-Mills theory}
\label{s.2+1}

In the previous sections we have introduced the notion of {\it volume independence}
and discussed how it arises in perturbation theory. We have shown that,  
with appropriate choices of the twist tensor,
physical observables depend on the combination $\tilde l = l N^{2/d}$. To check
whether this holds at a non-perturbative level, lattice simulations are
required. In the case of Yang-Mills theories in 2+1 dimensions, an exploratory analysis has 
been recently presented in Ref.~\cite{Perez:2013dra1}. It is the purpose
of this section to review part of those results.

Before doing that, let us recall what are the consequences of the perturbative 
analysis when particularized to SU($N$) Yang-Mills theories in 2+1 dimensions. 
A specific feature of three dimensions is the mass dimensionality of 't Hooft coupling.
When combined with the observation that perturbation theory at all orders depends
on $\tilde l$, this implies that all dimensionless quantities should depend on the 
variable 
\be
x={\lambda \lef \over  4\pi}
\ee
which thus becomes the relevant 
scale parameter \cite{Guralnik:2001pv},\cite{Perez:2013dra1}~\footnote{In four dimensions,
with one of them compactified, the authors of Refs. \cite{Unsal:2010qh1,Unsal:2010qh2} suggest that this role is played by 
$\lef \, \Lambda_{QCD} /(4 \pi)$.}. 
This is exemplified by the one-loop formula for the electric flux energy derived previously:
\be
\label{eq.oneloop}
{\ET^2(\vec e)  \over \lambda^2} =  { |\vec n|^2 \over 4 x^2}
- {1 \over   x } G\Big ({\vec{e} \over N}\Big )  \, ,
\ee
with $e_i =  \epsilon_{ij} n_j \, \kb $ (mod $N$) \footnote{We remind the reader that $\ttheta = 2 \pi \kb /N$,
 with $\kb$ defined through the relation: $k \kb = 1$ (mod $N$).}.
An interesting question is whether this $x$ dependence is preserved beyond perturbation theory. 
 Dimensionless 
quantities in the zero electric flux sector
should become volume and thus $x$ and $\ttheta$ independent in the large volume limit. Concerning
non-zero electric flux sectors, confinement predicts an energy of 
electric flux that rises linearly with the size of the box, leading to:
\be
{\ET(\vec e)  \over \lambda} =  {\sigma_{\vec e} \, l \over \lambda} \equiv 4 \pi x \,
{\sigma_{\vec e} \over N \lambda^2} \, .
\ee
If we parameterize the string tension for electric flux $\vec e$ as:
\be
\sigma_{\vec e}  = N \sigma \phi\Big( {\vec e \over N}\Big )\, ,
\ee
the relation between $\vec e$ and $ \vec n$ leads to a formula perfectly 
consistent with $x$-scaling for $\ttheta$ fixed:
\be
{\ET(\vec e)  \over \lambda} =  4 \pi x \, {\sigma \over \lambda^2} \, \phi \Big ( {\ttheta \vec{\tilde n}
\over 2\pi} \Big )\, ,
\ee
with $\tilde n_i = \epsilon_{ij} n_j$.
The function $\phi (z)$ encodes information on the scaling of the $k$-string tension with the
electric flux (or winding number of the $k$-string). 
The most common functions used for this purpose in the literature are:   
\be
\phi (z) = \sin(\pi z) /\pi \, ,
\ee
known as Sine scaling, and
\be
\phi (z) = z (1-z) \, ,
\ee
known as Casimir scaling. We will present below some results on $\phi(z)$ derived in 
Ref.~\cite{Perez:2013dra1}. By appropriately adjusting the value of $\ttheta$, 
one can explore the $z$-dependence of $\phi(z)$ for large values of $z$. This helps in providing 
stronger constraints on the type of scaling favoured by the data.  

One can also conjecture about the volume dependence of the energy of electric flux beyond the 
leading linear term.  The effective string description of the flux tube 
provides an expansion in terms of $1/(\sqrt{\sigma} \, l)$. It turns out that 
in 3 dimensions all terms up to order $1/l^5$ are universal~\cite{Luscher:2004ib,Aharony}, and agree with 
the ones derived from the Nambu-Goto string action.
Our set-up differs, however, from the standard one. The geometry is different since two
of the directions, instead of one, are compactified. In addition, they are twisted.
It has been suggested that the effect of the twist can be mimicked in 
the string description by the introduction of a Kalb-Ramond $B$-field background \cite{Guralnik:1999eb}. 
The observation that open strings have non-commutative gauge theories as a particular low energy 
limit~\cite{Seiberg:1999vs} makes this conjecture rather natural. Let us see how far we can push this analogy for closed 
strings.
The Nambu-Goto prediction for the energy of a closed
string winding $\vec e$ times around the torus on the background of a Kalb-Ramond $B$-field is given by:
\be
{\ET^2(\vec e)  \over \lambda^2} =  \Big ({ \sigma |\vec e| l \over \lambda}\Big )^2
- {\pi \sigma \over   3 \lambda^2 } + \sum_i \Big ({  \epsilon_{ij} e_j B \over \lambda l}\Big )^2  \, ,
\ee
The $B$-field is related to the non-commutativity parameter through:
$\theta_{ij} = - \epsilon_{ij} l^2 /B$. If we insert this relation in the Nambu-Goto expression, together with
Eq.~(\ref{eq.noncom}), we derive that the $B$-field contribution is identical to the perturbative tree-level 
term in the twisted box. This leads to an expression for the Nambu-Goto string given by:
\be
\label{eq.nambu}
{\ET^2(\vec e)  \over \lambda^2} =  \Big ({ 2 \sigma |\ttheta \vec{n}| \over \lambda^2}\Big )^2 \, x^2
- {\pi \sigma \over   3 \lambda^2 } +  {|\vec n|^2 \over 4 x^2}  \, ,
\ee
respecting $x$-scaling, at fixed $\ttheta$, also at this level. One interesting observation is that this formula combines in 
quadrature the first two terms in
the ordinary string description with the tree-level term of the perturbative expansion.  This suggests
a generalization of the form:
\be
\label{eq.etn}
{\ET^2(\vec e)  \over \lambda^2} =  \Big ( { 4 \pi  |\vec{e}| \sigma \over N \lambda^2}\Big )^2 \, x^2
 - {\pi \sigma \over   3 \lambda^2 } -  {1 \over   x } G \Big
({\vec e \over N} \Big ) +  {|\vec n|^2 \over 4 x^2}  \, .
\ee 
Notice that the confining term rises quadratically with $x$ and overcomes, at large values of $x$, 
the negative contributions from the self-energy and the constant term. A full discussion on the occurrence 
of tachyonic instability should thus take into account this non-perturbative contribution.

\subsection{Electric flux spectrum}
\label{s.results}

As already mentioned, the aim of this section 
is to test the prevalence of {\it volume independence} beyond perturbation theory. 
To achieve that purpose, we will review the outcome of a non-perturbative analysis in 2+1 dimensions 
carried out in Ref.~\cite{Perez:2013dra1}.  We will not describe the results in full detail 
but will instead single out those that allow to test if and when tachyonic instabilities occur.

Let us start the presentation with a brief description of the numerical set-up. Space and time have been discretized on a 
$N_s^2 \times N_0$ lattice. We have employed the Wilson plaquette action modified to take into account 
the twisted boundary conditions.
The procedure is standard and amounts to introduce a plaquette twist-dependent factor. With 
this the lattice action reads:
\be
S_W = N b \sum_{n \in {\bf Z}^3}  \sum_{\mu \ne \nu} \Tr \Big \{ 1\!\!\!1  - z_{\mu \nu}^*(n) \, 
U_\mu (n) U_\nu(n+\mu) U_\mu^\dagger (n+\nu) U_\nu^\dagger (n)  \Big \} \quad ,
\label{eq.wilson2}
\ee
with $U_\mu$ the SU($N$) link matrices, and 
where $z_{\mu \nu}(n)$ is equal to 1 except for the corner plaquettes in each (1,2)
plane where it takes the value:
\be
z_{i j}(n) = \exp  \Big \{i\,  {2\pi \epsilon_{ij} k \over N} \Big \}  \quad .
\ee
The quantity $b$ is proportional to the inverse of the dimensionless lattice 't Hooft coupling: 
$b \equiv 1/(a \llat)$, with $a$ the lattice spacing.
Exploring {\it volume independence} requires to perform lattice simulations at various values of
 $k$, the gauge group SU($N$), and the physical size of the torus $l$. 
The study in Ref. \cite{Perez:2013dra1} is an exploratory one, trying to address some of the main
concerns raised in sec.~\ref{s.beyond}. We have selected for that purpose a set of $N$ and $N_s$
values that give an approximately constant value of $N N_s$. By varying $b$ one can thus cover a 
wide range of values of the variable $x$ which, in terms of the lattice quantities, 
reads:  $x_L = N N_s / (4 \pi b)$. A full continuum extrapolation of the results has not been 
attempted yet. In the coarsest lattices that have been analyzed, we have observed a mild lattice spacing 
dependence of the electric flux energies, but it does not alter the main conclusions of 
the analysis that will be presented here. The reader interested in having further details 
concerning the simulations and a full account of results should consult Ref. \cite{Perez:2013dra1}.

The discussion will be restricted to the sectors of electric flux that are generated from
straight line Polyakov loops winding $e$ times along the torus. They are projected over 
the minimal momentum in each electric flux sector $\vec p = (2\pi n /\lef, 0)$, with 
$n= k e$ (mod $N$). To simplify the notation, the corresponding energies will be denoted 
by $\ET_n$. Numerically they have been extracted from the exponential decay at large times of  
spatially smeared Polyakov loop correlators. 

\begin{figure}
\centerline{
\psfig{file=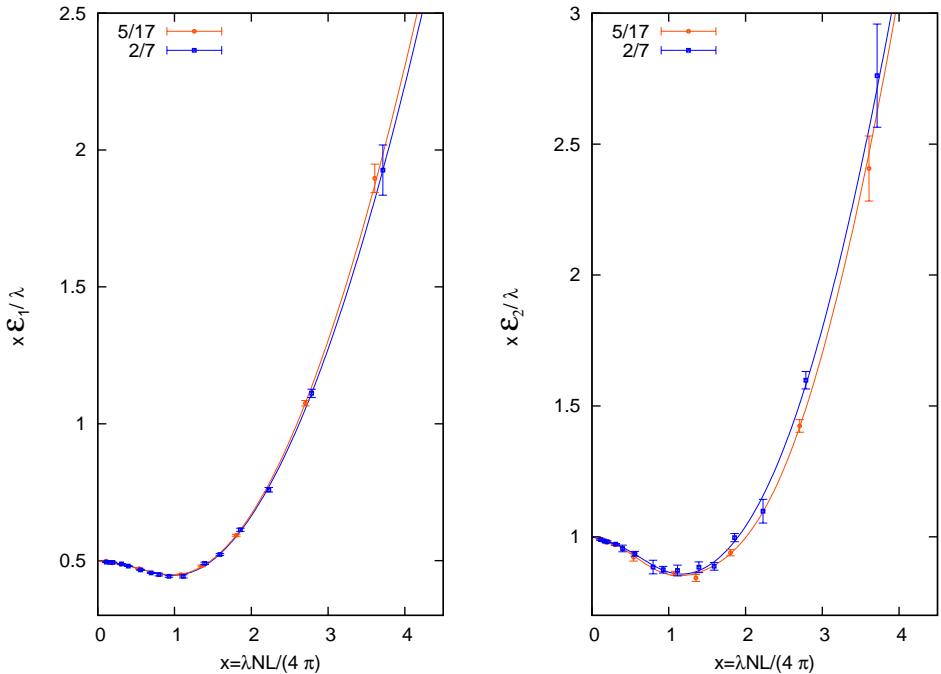,angle=-90,width=13cm}}
\caption{
We display $x\ET_n /\lambda$, with $n=1,2$, as a function of $x=  \lambda  \lef /(4\pi)$.
The results correspond to gauge groups SU(7) and SU(17) with the values of $k$ adjusted to obtained
approximately equal values of $\ttheta /(2\pi) = \kb /N$, indicated in the plot.}
\label{fig.etn}
\end{figure}
\begin{figure}
\centerline{
\psfig{file=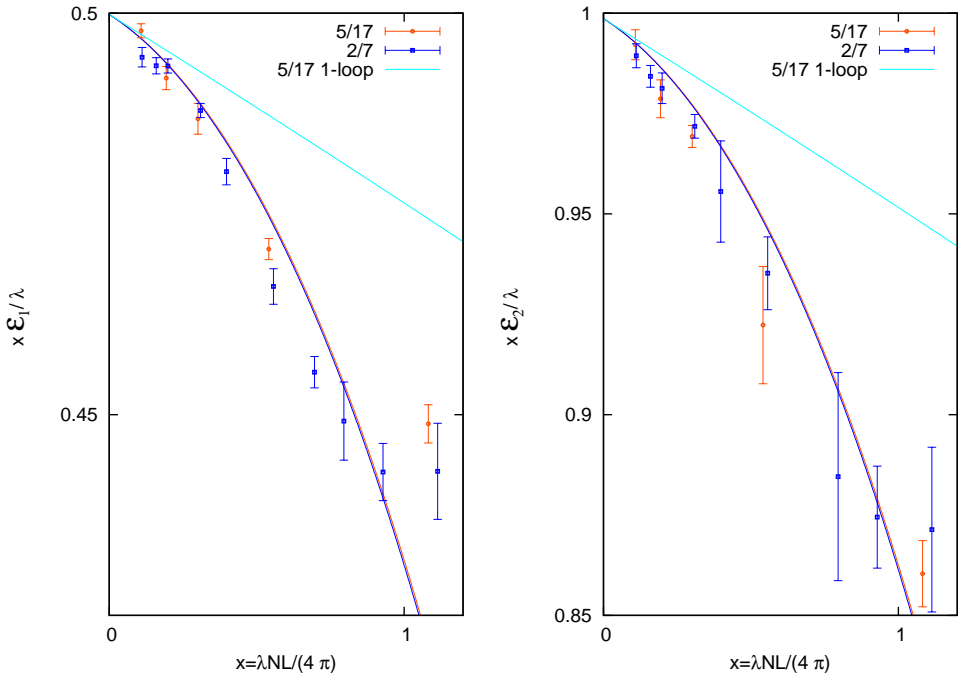,angle=-90,width=13cm}}
\caption{
The same as in fig.~\ref{fig.etn} with the low $x$ region enhanced.
The line denoted as {\it 1-loop} corresponds to the one-loop prediction
Eq.~(\ref{eq.oneloop}). The others correspond to Eq.~(\ref{eq.thres}) with $n$ fixed either to
1 or 2.  }
\label{fig.etn.small}
\end{figure}

We will first focus on values of $k$ and $\kb$ that satisfy the conditions imposed 
in sec.~\ref{s.inst} to prevent the occurrence of instabilities. A comparison will be made 
between SU(7) and SU(17) at very close
values of $\ttheta$ equal to $2/7$ and $5/17$, respectively.  
The spectrum is classified by the values of $(n,\ttheta n)$.
The energies, multiplied by $x$ and corresponding to $n=1$ and $2$, are displayed as 
a function of $x$ in fig.~\ref{fig.etn}.  They show a universal scaling  with $x$, irrespective of the value of $N$.
The general features shown in the figure are also in good correspondence with our expectations. 
The energies start at $|\vec n|/2$, the tree-level perturbative result, and decrease
as $x$ increases due to the self-energy contribution.

For the moment, let us discuss the small $x$ region. The large $x$ confinement regime will
be addressed later on. An enlarged version of the plots, singularizing the small $x$ dependence, 
is shown in fig.~\ref{fig.etn.small}. The one-loop prediction is followed for very low $x$. Notice, 
however, that the range of $x$ values displayed is too large to rely solely on perturbation theory.
A surprisingly good description of both $\ET_1$ and $\ET_2$ is provided instead by the formula:  
\be
\label{eq.thres}
{x \ET_n  \over \lambda} =
|\vec n|\,  \Big( \, {1 \over 4 } - x \, G 
(\ttheta/ 2\pi ) - {\pi \sigma x^2 \over   3 \lambda^2 }\,  \Big )^\half  \, ,
\ee
where we have fixed $\sqrt{\sigma}=0.19638(9)$, the value of the 2+1 SU($\infty$) string tension
determined by Teper and collaborators in \cite{teper1}-\cite{teper7}. This expression contains, in addition to the 
perturbative result, 
the first relevant term in the Nambu-Goto string expression Eq.~(\ref{eq.etn}), which gives also 
a negative contribution to the energy squared. 
With no free-parameters this equation describes quite well the data up to values of $x\sim 1$.
Notice, however, that for $x>1$ the energies displayed in Fig.~\ref{fig.etn} start to grow, as predicted
by confinement, and do never become zero. 
This shows, as anticipated, that the perturbative formulas cannot be trusted when 
$x$ becomes of order 1, and that instability does not occur in these data sets.

\begin{figure}
\centerline{
\psfig{file=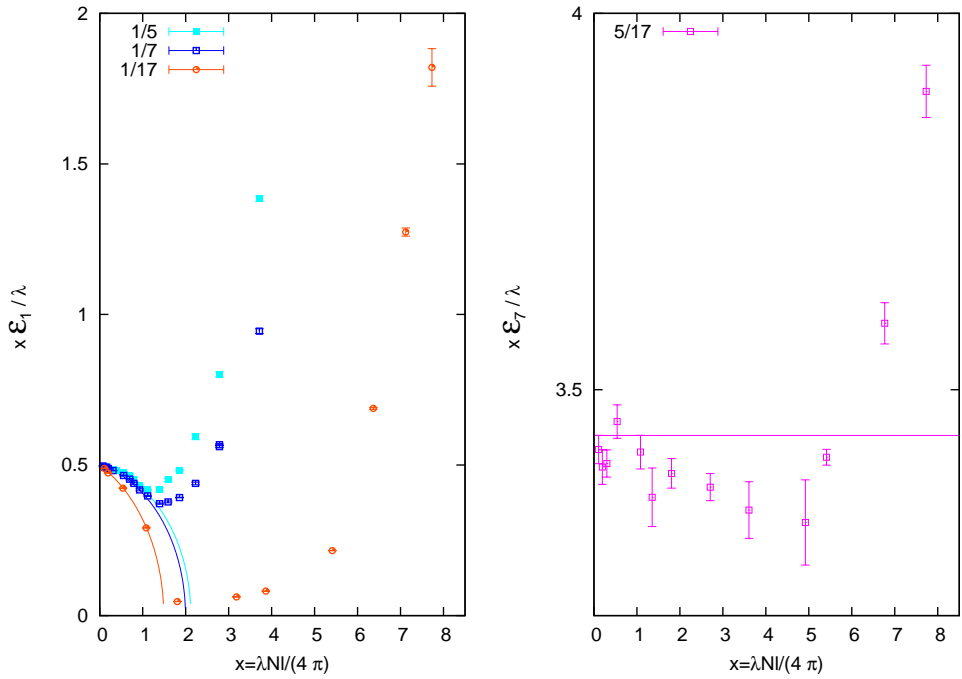,angle=-90,width=13cm}}
\caption{We display $x\ET_n /\lambda$, as a function of $x= \lef \lambda /(4\pi)$, for states with electric
flux one. The plot on the left corresponds to $\ttheta/(2\pi) = 1/5,1/7,1/17$. The lines in the plot correspond to
Eq.~(\ref{eq.thres}) for momentum $n=1$.
The plot on the right corresponds to $\ttheta/(2\pi) = 5/17$.
The line in the plot corresponds to the tree-level lattice value for $n=7$.}
\label{fig.etn.inst}
\end{figure}

For a comparison, we have also examined what happens if $\ttheta$ tends to zero in the large $N$ limit.
Figure~\ref{fig.etn.inst}  shows the results corresponding to the lowest energy state for
$\ttheta/(2\pi)=1/5,1/7,1/17$. It corresponds to the state of electric flux $e=1$ with momentum
$n=1$.  For $N=17$, $\ET_1$ gets very close to zero at $x_c\sim 1.5$ and stays very low in a window 
of intermediate values of $x$, until the confinement term starts to dominate and reverts this behaviour.
Although this situation only takes place at intermediate values of $x$, one expects that
in the $N\rightarrow \infty $ limit this regime would extend over the full $x$-axis.
In Ref.~\cite{Guralnik:2002ru} it has been conjectured that this situation corresponds to a phase in which
electric flux condenses and the Polyakov loops acquire a vacuum expectation value.
One could think that something similar would take place for small values of the electric flux irrespective
of the value of $\ttheta$ and $k$. 
To see that this is not the case, let us look for instance at SU(17) 
with $\ttheta=5/17$ and $k=7$. The momentum corresponding to $e=1$ is $n=7$. 
The $x$-dependence of $x\ET_7/ \lambda$ is displayed in the right plot of fig.~\ref{fig.etn.inst}. 
For this value of $n$, the tree-level
term is sufficiently large to push the threshold of instability to the region where confinement
is already relevant, therefore avoiding the occurrence of instability.

\begin{figure}
\centerline{
\psfig{file=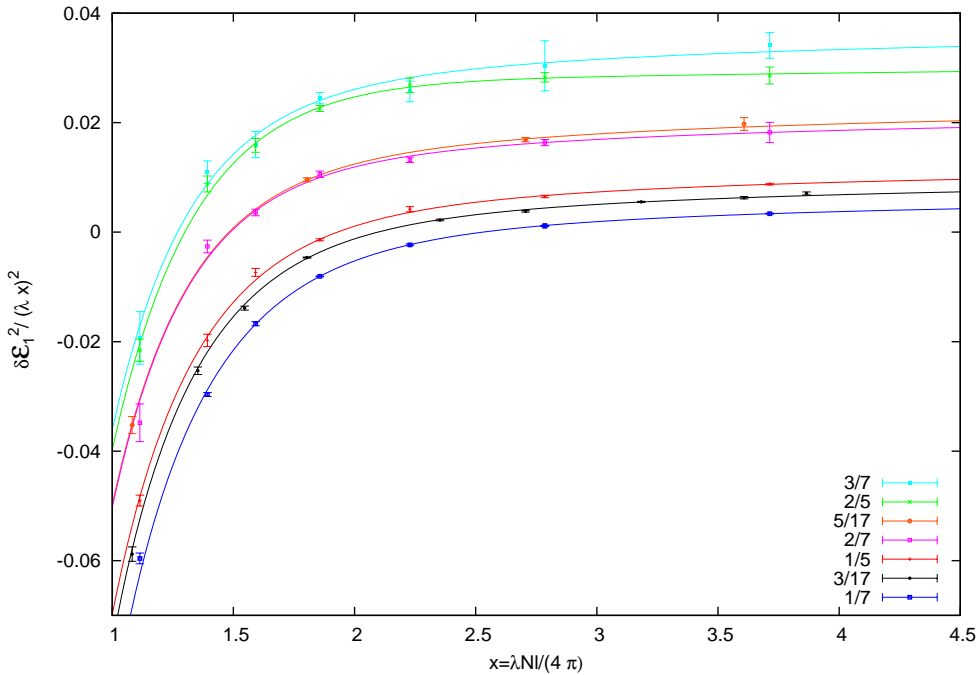,angle=-90,width=13cm}}
\caption{\label{fig.deltaE2}
We display $\ET_1^2 /(\lambda x)^2$ after subtracting out $1/(4 x^4)$, which is the zero-order
perturbative contribution to this quantity. The lines are fits to Eq.~(\ref{eq.etn2}). 
The labels in the plot indicate the value of $\ttheta /(2\pi) \equiv \kb /N$.
}
\end{figure}

\begin{figure}
\centerline{
\psfig{file=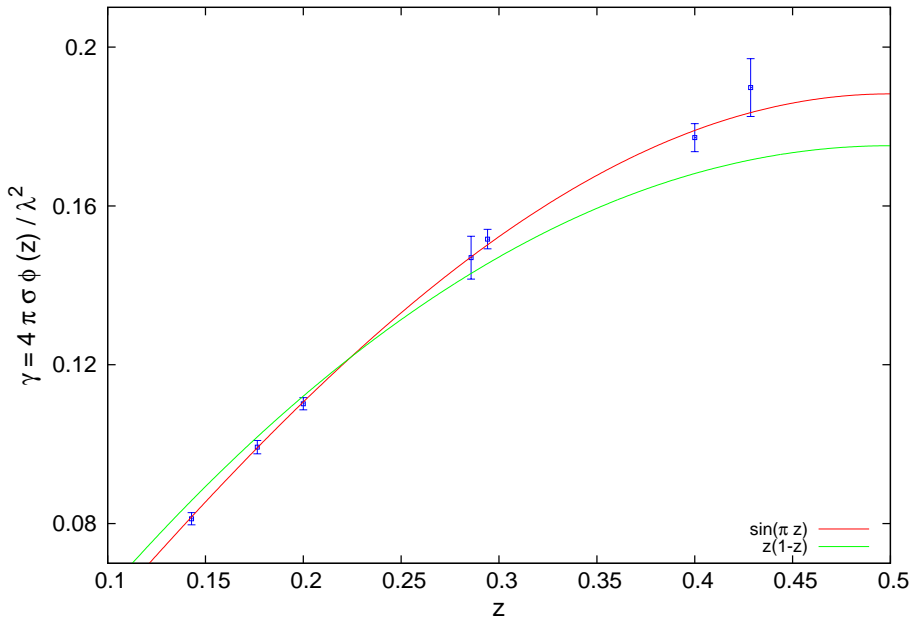,angle=-90,width=12cm}
}
\caption{\label{fig.kstring}
We display the function
$\gamma(z) = 4 \pi  \sigma  \phi(z)/ \lambda^2$, given  in Eq.~(\ref{eq.gamma}), and representing
the electric flux dependence of the $k$-string tension ($z=e/N$).
The red line in the plot is a fit to the Sine scaling formula: $\phi(z) = \sin(\pi z) /\pi$.
The green line corresponds to the prediction from Casimir scaling: $\phi(z) = z(1-z)$.
}
\end{figure}

We come now to the analysis of the large $x$ regime. We will restrict our attention to the study of $\ET_1$
for various values of $\ttheta$. This will allow us to investigate the dependence of the string tension on the 
electric flux going up to larger values than those previously studied in the literature.
The values of $(\ET_1)^2/(\lambda x)^2$ after
subtracting out $1/(4x^4)$, which is the zero-order
perturbative result for this quantity, are displayed in fig.~\ref{fig.deltaE2}.
The curves should tend at infinity to the string tension, approaching this limit with
a $1/x^2$ dependence. This is consistent with the observed behaviour, with the order or
the curves reflecting the order of the values of $e/N = \ttheta/(2\pi)$. 
One would like to extract from these curves a prediction for $\phi(z)$, the function
giving the dependence of the string tension on the electric flux:
\be
\sigma_{\vec e}  = N \sigma \phi\Big( {\vec e \over N}\Big )\, ,
\ee
A quantitative analysis requires to fit the $x$ dependence of the curves in order to extract the asymptotic
value. In Ref.~\cite{Perez:2013dra1} we proposed a fitting function based on Eq.~(\ref{eq.etn}) which describes very well all 
the data, if one allows for the addition of one extra term inspired by the form of instanton-like contributions. 
The final fitting function, corresponding to the lines displayed in fig.~\ref{fig.deltaE2}, is of the form:
\be
\label{eq.etn2}
{\delta \ET^2_1(z)  \over (\lambda x)^2}  \equiv {\ET^2_1(z)  \over (\lambda x)^2} - {1 \over 4 x^4} =  \gamma^2(z)  - {\gamma(z) \over   12 x^2 z (1-z)} 
+ {{\cal A}(z) \over x^5 \sqrt{x}}  \, e^ {- {S_0(z) \over x}} -  {G(z) \over   x^3 } 
  \, ,
\ee
with $z=\ttheta/(2\pi)$. We will not provide the details of the fitting procedure here, the interested reader can 
consult them in Ref.~\cite{Perez:2013dra1}. The results for $\gamma(z)$ allow us to study the electric flux
 dependence of the string tension. We parameterize it as:
\be
\label{eq.gamma}
\gamma(z) = {4 \pi \sigma \over \lambda^2} \phi(z)
\ee
In fig.~\ref{fig.kstring}  we display $\gamma(z)$ for different values of $\ttheta$.
The lines displayed correspond to Casimir scaling: $\phi(z) = z(1 - z)$, and the Sine scaling: $\phi(z) = \sin(\pi z)/z$.
The fit corresponding to Sine scaling is clearly much better, giving 
a $\chi^2$ per degree of freedom of 0.26. The value extracted from the fit for the fundamental string tension is 
$\sqrt{\sigma}/\lambda = 0.217(1)$, deviating around 10$\%$ from the value obtained by Teper and collaborators in Ref.~\cite{teper1}.
Given the absence of a continuum extrapolation in our data, the agreement can be considered very satisfactory.

\section{Conclusions}
\label{s.con}

In this review we have discussed the idea of {\it volume independence} introduced in
Ref.~\cite{Perez:2013dra1}. This notion arises naturally when dealing with SU($N$)
Yang-Mills theories defined on even-dimensional tori endowed with twisted boundary conditions.  
Its first obvious manifestation is the fact that the perturbative
series, to all orders in 't Hooft coupling, depends jointly on a combination of the 
rank of the group ($N$) and the periods of the torus ($l$), given  by $\tilde l = l N^{2/d}$.  
This holds for irreducible twist tensors $n_{\mu\nu} =\epsilon_{\mu \nu} \, k  N / L $, with
$k$ and $L\equiv N^{2/d}$ coprime integers. 

The precise statement is that
all vertices in perturbation theory are proportional to the factor:
\be
\sqrt{\frac{2\lambda} {\prod_\mu \lef_\mu}}\, \, \sin\Big (\frac{\theta_{\mu \nu}}{2} 
\, p_\mu  q_\nu\Big ) \, ,
\ee
where
\be
\theta_{\mu \nu} = \frac{ \lef_\mu \,  \lef_\nu} {4\pi^2} \times  \, \tilde
\epsilon_{\mu \nu} \, \tilde \theta  \, ,
\ee
with $\ttheta= 2 \pi \bar k/N$ ($\kb$ depends on $k$ and is defined by $k \kb = 1$ 
(mod $L$)). If all physical quantities depend smoothly on $\ttheta$, this implies an equivalence 
between different SU(N) Yang-Mills theories defined at fixed values of $\ttheta$ and $\lef$. 

This idea links in a natural way to the old proposal of Eguchi Kawai reduction extending its
validity to finite values of $N$. Indeed our description follows closely the 
derivation of reduction presented for the Twisted Eguchi Kawai model in 
Refs.~\cite{TEK12,EguchiNakayama} and, in particular, the derivation of the 
momentum dependent Feynman rules that were the precursors of non-commutative field theory 
\cite{GAKA}. In order to make the review self-contained, we have discussed in detail the 
perturbative set-up, as well as the connection to TEK and non-commutative gauge theory.

In the rest of the paper, we have addressed the question of whether {\it volume independence} 
holds beyond perturbation theory. We have started by discussing possible caveats, including
the occurrence of tachyonic instabilities at one-loop order~\cite{Guralnik:2002ru}, 
or the breaking of translation symmetry due to non-perturbative effects in TEK reduction
\cite{IO}-\cite{Azeyanagi}.
We have argued that a judicious choice of $k$ and $\kb$, as the one
advocated in ~\cite{TEK21}, is sufficient to avoid both problems.

Resolving the non-perturbative fate of {\it volume reduction} requires, however, 
to perform
numerical simulations. In the last part of the paper we have presented the results of an exploratory
analysis of these issues in 2+1 dimensions \cite{Perez:2013dra1}. For this particular case, the 
predicted $\lef$ dependence merges with the (dimensionful) 't Hooft coupling dependence in
the variable $x= \lambda \lef /(4\pi)$. For fixed $\ttheta$, volume independence then amounts 
to a universal scaling in $x$ of all dimensionless physical quantities. We have tested this idea by 
analyzing the $x$-dependence of the electric flux energies. Remarkably, the theoretical 
expectations for the large volume confining regime satisfy $x$-scaling and allow us to extract
information about the electric flux dependence of the $k$-string tension.

The main focus of the numerical analysis presented here, has been though to test the conditions 
under which tachyonic instabilities are absent. We have presented indications
that, for certain values of $k$, the intermediate volume regime might indeed
be affected by the instability. Nevertheless, we have also shown cases where this is avoided if the twist is 
selected according to the criteria reviewed in sec.~\ref{s.beyond}. 
 
Let us finally mention that an important test of {\it volume independence} would be to address 
the $\lef$ dependence of the zero-electric flux sector, and in particular of the 
glueball spectrum. In the large volume regime, these quantities should become 
independent of the boundary conditions and hence of $\ttheta$.  This should also hold in the
large $N$ limit, irrespective of volume effects, if {\it volume independence} is preserved
non-perturbatively. We have at present an ongoing project that will address in detail all 
these issues.

\section*{Acknowledgments}

On the sad occasion of the death of our dear friend and colleague Pierre van Baal, we would like to 
dedicate this review to honour his memory. Pierre's many contributions have left a profound imprint in 
our present understanding of the femto-universe, and have strongly influenced 
the field. His masterworks have been recently collected in the book  
``Taming the forces between quarks and gluons - Calorons out of the box - Scientific papers by Pierre
van Baal" ~\cite{vanbaal:book}. They reflect the passion Pierre had for scientific challenges and 
his impressive talent for analytic calculations. One of us, MGP, would like to express 
her deep gratitude to Pierre for countless discussions over the years and for the burst of ideas and enthusiasm
that he so generously injected in all his collaborations.

M.G.P. and A.G-A acknowledge financial support from the grants FPA2012-31686 and 
FPA2012-31880,  the MINECO Centro de Excelencia Severo Ochoa Program SEV-2012-0249,
the Comunidad Aut\'onoma de Madrid  HEPHACOS S2009/ESP-1473, and the EU 
PITN-GA-2009-238353 (STRONGnet). They participate in the Consolider-Ingenio 
2010 CPAN (CSD2007-00042). M. O. is supported by Grants-in-Aid for Scientific Research from the 
Japanese Ministry of Education, Culture, Sports, Science and Technology (No 26400249). 
We also acknowledge the use of the IFT HPC-clusters for the numerical simulations presented in this review.

\end{document}